\documentclass[9pt,twocolumn,twoside]{osajnl}
\journal{ol} % Choose journal (ao, aop, josaa, josab, ol, optica, pr)

\setboolean{shortarticle}{true}
\title{Single-Photon Vibrometry}
\author[1,2]{Patrick Rehain}
\author[1,2]{Jeevanandha Ramanathan}
\author[1,2]{Yong Meng Sua}
\author[1,2]{Shenyu Zhu}
\author[1,2]{Daniel Tafone}
\author[1,2,*]{Yu-Ping Huang}

\affil[1]{Department of Physics, Stevens Institute of Technology, 1 Castle Point Terrace, Hoboken, NJ, 07030, USA}
\affil[2]{Center for Quantum Science and Engineering, Stevens Institute of Technology, 1 Castle Point Terrace, Hoboken, NJ, 07030, USA}

\affil[*]{Corresponding author: yuping.huang@stevens.edu}

\begin{abstract}
We propose and demonstrate a single-photon sensitive technique for optical vibrometry. It uses high speed photon counting to sample the modulated backscattering from a vibrating target. Designed for remote vibration sensing with ultralow photon flux, we show that this technique can detect small displacements down to 110 nm and resolve vibration frequencies from DC up to several kilohertz, with less than 0.01 detected photons per pulse. This single-photon sensitive optical vibrometry may find important applications in acousto-optic sensing and imaging, especially in photon-starved environments.
\end{abstract}

\setboolean{displaycopyright}{true}

\begin{document}

\maketitle
Optical means have become the tool of choice for the measurement and profiling of mechanical vibrations \cite{ROTHBERG201711,CASTELLINI20061265,doi:10.1063/5.0004363}. In optical vibrometry, a portion of the returning optical probe is detected and the target's displacement or vibration information is extracted from the measured signal. To this end, interference based detection mixes the returning signal with a reference beam and demodulates the target motion from the beating between the two. This method uses the phase coherence between the probe and reference to achieve displacement sensitivity on the order of a fraction of the optical wavelength used \cite{doi:10.1063/5.0004363}. While such interferometry offers high performance in terms of bandwidth and displacement sensitivity, it requires a laser with long coherence time and strong and stable power in the returning beam for faithful measurement. These requirements preclude the interference-based systems from operating when the signal is very weak, and makes them sensitive to intensity fluctuations arising from the dynamic speckle patterns returning from a rough target. 

The speckle noise present in interference-based vibrometry manifests as an intensity modulation that distorts the desired phase-modulated signal. Demodulating the two effects can be exceedingly difficult because they occur at the same frequency. Speckle noise has been widely studied and various techniques have been shown to mitigate its effects in certain applications \cite{LV2019117,Rothberg:06,MARTARELLI20062277}. For example, it can be detected using the kurtosis ratio of the measured electronic signal and then reduced via a combination of both time and frequency domain filtering \cite{VASS2008647}. Also, a scanning average technique can blur the speckle-pattern by rapidly sweeping the beam across the target surface and later dropping the scanning frequency \cite{Zhu:19}. In another method, a balanced detection setup using two photodiodes is shown to be highly resilient against intensity fluctuations by taking a differential measurement between two phase angles \cite{Rzasa:15}. These are just a few example of techniques that can mitigate the speckle noise, and while they can be effective, they can be costly and complicated to implement. The need for a strong optical signal and the distortions caused by speckle noise remain the major limiting factors in the interference-based vibrometry. 

%The challenge presented by speckle noise in interference-based vibrometry amounts to isolating the desired phase modulated signal from the intensity modulation caused by changing speckle patterns. 

Recognizing these limitations, intensity-based vibrometry has been used to probe target vibrations by directly measuring the speckle dynamics. This has been demonstrated using a CCD-camera to track the motion of individual speckles using post-processing algorithms \cite{Bianchi:14, Zalevsky:09}. These systems are limited by the camera frame rate, the distribution of the intensity across many pixels, and the computational cost of processing all pixels. Recently, a modified version of this setup was introduced where Newtonian telescope was used to focus the 2D speckle pattern onto a line-scan CCD \cite{Bianchi:19}. This increased the maximum frame rate and the intensity at each pixel, while reducing the dimensionality of post processing. Single-pixel systems have also been used by directly reading the intensity-modulated output of a photodiode\cite{Bianchi:14,Veber}.  An aperture or mask is placed in front of the detector to tune the number of collected speckles. The average intensity is increased by summing over more speckles, but comes at the cost of signal contrast (modulation depth). The single photodiode can achieve much higher frame rate and detection sensitivity than a CCD camera, but is subject to harmonic distortion at large amplitudes. While these intensity-based methods can increase the detection SNR, they cannot operate under highly photon-starved conditions when the returning signal is single-photon level (below the sensitivity of a photodiode).

Pushing the limits on remote sensing involves practical scenarios where the returning optical signal is very weak. While in principle this can be overcome by increasing the output laser power and detection dwell-time, this is not always possible in practice. High output power can lead to eye-safety concerns, meaning the laser power cannot be arbitrarily increased. Time correlated single photon counting (TCSPC) techniques using pulsed laser source synchronized to a Geiger-mode avalanche photodiode (APD) have proved successful in 3D imaging systems with less than one detected photon per pulse \cite{Pawlikowska:17,2018NatSR...817726H}. These systems accumulate detections over long dwell-times and use photon counting statistics to extract the target information. In vibration sensing, high-speed data acquisition is a necessity as the detection bandwidth is inversely related to the dwell-time.

In this letter, we demonstrate a single-photon sensitive technique for remote displacement and vibration sensing, based on direct photon counting the ultra-weak (single-photon level) backscattered optical signal by using a Geiger mode APD. Our single-photon vibrometry (SPV) technique measures changes in the photon flux that occur as the target surface is slightly displaced and tilted, similar to other intensity based systems, but is suitable for low flux applications where the number of returning photons is much less than one per probe pulse. Using 50 MHz repetition rate pulses, the variation in photon counts is captured by integrating the number of photon detections over a preset dwell-time ($\Delta t$) using a field-programmable-gate-array (FPGA). This process is repeated to form a time-series of photon counting measurements with sampling rate limited by the programmable bandwidth set by the FPGA. The digital nature of photon counting circumvents the need for a complex chain of electronic amplifiers and filters commonly found in analog detection systems. The time-series measurement generated by the FPGA is passed via ethernet to a computer for analysis. 

When a vibrating target is well sampled with high contrast to noise ratio, the motion of the target is directly manifest in the time-series. The vibration frequency can be resolved even when the signal contrast is below the shot noise by identifying the peaks in the Fourier Transformed spectrum. SPV is fundamentally limited by the Poisson noise that is intrinsic to any photon counting measurement. Nonetheless, the statistically random nature of such shot noise leads to unbiased baselines in both time and frequency domains, which can be mitigated through post-processing. While we only show a simple method of post-processing in this paper, the one-dimensional data offer an easily accessible platform for more extensive digital signal processing techniques such as signal smoothing, frequency filtering, and harmonic analysis. We show SPV can sense mechanical displacements down to 110 nm, and resolve vibration frequencies from DC up to 4 kHz with less than 0.01 detections per pulse.

%we show that it is possible sense sub-micron mechanical displacements, and resolve vibration frequencies from DC up to several kHz, in photon-starved environments, by directly detecting the time varying optical intensity of a pulsed laser source using a Geiger mode APD. Our single photon vibrometer (SPV) operates in low-flux regime where the number of detected photons is much less than one per probe pulse. In this regime, the returning intensity is measured by counting the number of detections (N) over a fixed dwell-time ($\Delta t$). 

%The pulsed source increases the likelyhood of detecting the weak return over background noise. The time-gating of the APD provides inherent ranging information by measuring the time-of-flight (ToF) of the pulses, and enables single-photon sensitive detection.

%This combination of source and detector is commonly used in photon-starved applications such as ranging and 3D imaging, but until now has not been demonstrated for remote vibration sensing.

%The pulsed source increases the likelyhood of detecting the weak return over background noise. The time-gating of the APD provides inherent ranging information by measuring the time-of-flight (ToF) of the pulses, and enables single-photon sensitive detection. 
\begin{figure}[htbp]
\centering
\fbox{\includegraphics[width=\linewidth]{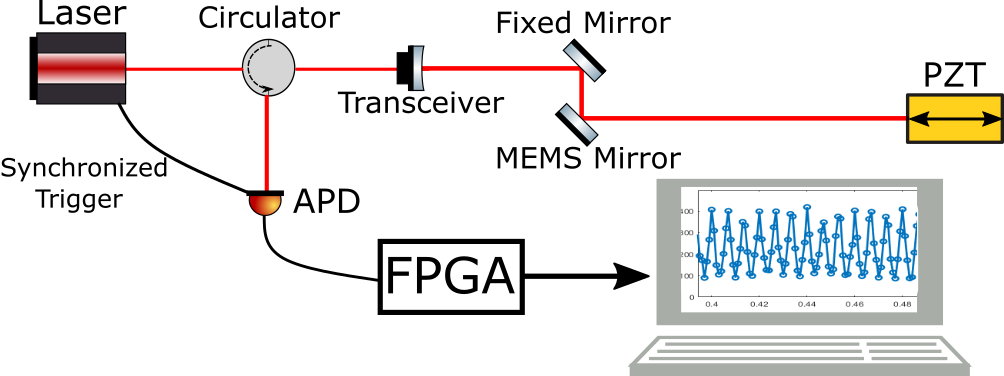}}
\caption{Experimental setup. A Laser sends an optical pulse train to be launched from a transceiver, and a synchronizing electronic signal to trigger the time-gate of InGaAs APD. A probe beam is guided towards PZT using mirrors. A fiber circulator passes the returning probe to APD for detection.}
\label{fig:setup}
\end{figure}

The current SPV setup is shown in Fig \ref{fig:setup}. A mode-lock laser
is used to generate the pulse train with 6 ps full width at half
maximum (FWHM) and 50 MHz repetition rate at 1554 nm. The laser sends a synchronized electronic pulse to the InGaAs APD to trigger the time-gated detection. Collimated probe pulses (Gaussian beam diameter: 2.2 mm) are transmitted toward the scene through an optical transceiver and programmable scanning MEMS mirror.  A fiber optics circulator separates the outgoing signal pulses and the incoming backscattered photons with a minimum isolation ratio of 55 dB. The transceiver is based on a simple monostatic coaxial arrangement using off-the-shelf optical components. It is comprised of an angle-polished single mode fiber (SMF) coupled to an aspheric lens, providing diffraction limited collimation of the probe laser beam. The MEMS mirror allows for easy beam steering towards the desired target, and the use of a coaxial transceiver negates the need for careful alignment between separate emitter and receiver when the distance to the target is changed. The low numerical aperture of the SMF acts as a spatial filter, with a small angle of acceptance, making the transceiver sensitive to spatial and angular changes in the returning signal that occur as the target moves. Intensity-based vibration sensors using the coupling sensitivity of SMF have been demonstrated previously \cite{Garca2010VibrationDU,4717237,CAO2007580,YANG2014333}, but these sensors are built for very short working distances for localized displacement sensing and do not achieve single-photon sensitivity.

We test the performance of SPV using a piezo-electric-transducer (PZT) placed 1 meter from the transceiver, for precise control over the target motion. The beam was fixed at a nearly normal angle of incidence to the surface of the PZT, and parallel to the axis of displacement. The average outgoing power was 0.15 mW and additional fiber attenuators were used to reduce the strength of the returning signal prior to detection, thus imposing photon-starved conditions. Direct detection of the attenuated signal is done using an InGaAs APD with detection efficiency of 10 \%  within a 1 ns gate.  The detector dark count rate is about 50 kHz and the deadtime is 100 ns.   

\begin{figure}[htbp]
\centering
\fbox{\includegraphics[width=\linewidth]{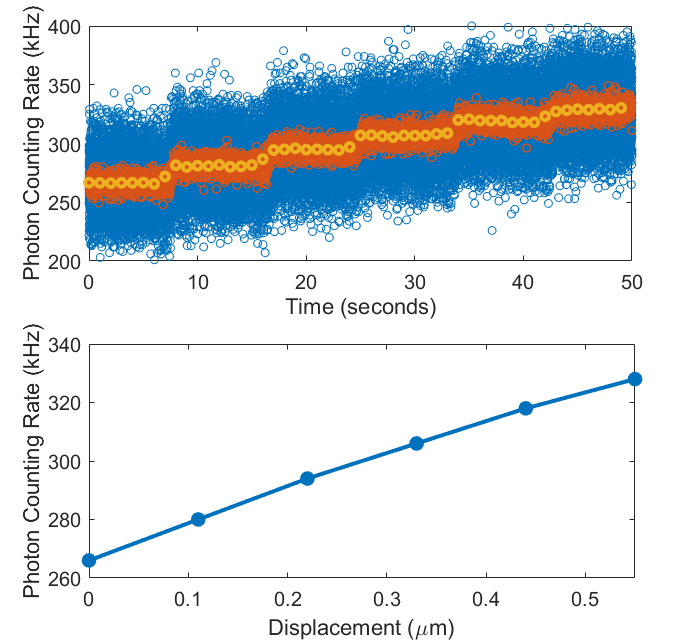}}
\caption{Top: Photon counting rate time-series with dwell-times of 1 ms (blue), 10 ms (red), and 1 second (yellow). Bottom: Photon counting rate as a function of the PZT displacement.}
\label{fig:displacement}
\end{figure}
 
Data acquisition is done using a FPGA to repeatedly count the number of detections (N) over a fixed dwell-time ($\Delta t$). The results are saved to a file and then sent to a computer for processing. The time varying photon count is given by
\begin{equation}
\label{eqn1}
N(t) = \Phi(t)\eta_{det} \Delta t,
\end{equation}
where $\Phi (t)$ is the collected photon flux (photons/second), and $\eta_{det}$ is the detector efficiency. Note $\Phi(t)$ is directly proportional to the repetition rate of the laser and Eq.~(\ref{eqn1}) holds when the photon detection rate is much less than 1 per gate. Considering the Poisson nature of photon counting, the detection SNR is given by $\sqrt{N}$, and scales as $\sqrt{\Delta t}$. Better SNR can be obtained with longer dwell-time, but comes at the cost of detection bandwidth ($\Delta B = 1/2\Delta t$). Following the Nyquist theorem, the maximum resolvable frequency is half the sampling rate. Target displacement is detected by a differential in photon counting $\Delta N = |N_1 - N_2|$, where $\Delta N$ is the signal contrast. The contrast to noise ratio (CNR) is defined as $\Delta N/(\sqrt{N_1 + N_2})$ \cite{BROWN1992,Pellegrini_2000,Shahverdi:18}. When the contrast is low, we can approximate this using the mean  ($\bar{N})$. The condition for discriminating between two measurements amounts to having CNR greater than 1, or $\Delta N >  \sqrt{2\bar{N}}$.

\begin{figure}[htbp]
\centering
\fbox{\includegraphics[width=\linewidth]{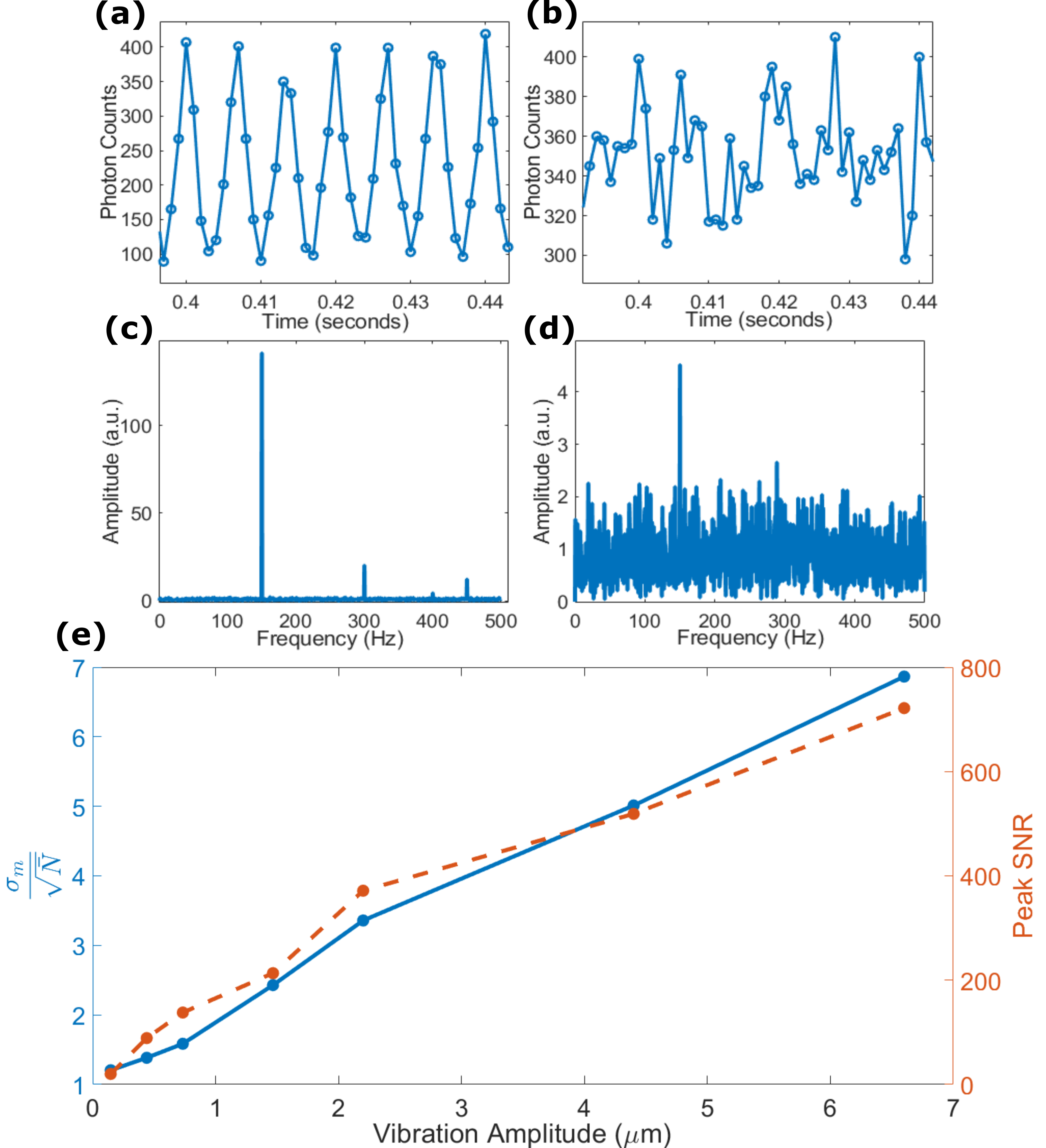}}
\caption{(a) Time-series of 150 Hz signal with 6.5 micron amplitude, where the contrast created is high enough that we can clearly trace the motion in the time-domain. (b) The same setup but with 0.15 micron amplitude, where the Poisson noise distorts time-domain signal. (c) FFT of (a), showing peaks at first three harmonics. (d) FFT of (b), still resolving driving frequency over noise, but not showing 2nd and 3rd harmonics. (e) Blue solid line is the ratio of the measured standard deviation and shot noise in the time-domain. Orange dashed line is the peak SNR in the frequency domain.}
\label{fig:amplitude}
\end{figure}

We first demonstrate the performance of this system in detecting small displacements with low signal flux. A singular displacement leads to a shift in the mean counting rate, this amounts to a change in the DC level of the signal. Figure \ref{fig:displacement} (top) shows the time-series acquired with dwell-times of 1 ms (blue), 10 ms (red), and 1 second (yellow). In each case the PZT is displaced 0.55 $\mu m$, in steps of 0.11 $\mu m$. The average counting rate was 275 kHz, meaning only 0.0055 detections per pulse. The signals in the figure are normalized by the dwell-time, making the contrast independent of the dwell-time, and allowing for easier visualization. As seen, the PZT stepping shifts the mean counting rate, and the shorter dwell-times lead to more overlap between the signal levels, making them harder to distinguish. Figure \ref{fig:displacement} (bottom) shows the linear change in the counting rate, with the contrast equal to about 15 kHz per 0.11 $\mu$m step. The counting rates are obtained from the 1 second dwell-time results, where there is almost no deviation in the signal at each level. The detection noise level of the 1 ms dwell-time (blue) is 22 kHz, measured by taking the standard deviation of a one second segment. Given the mean counting rate of 275 kHz, the detection noise level is about 5 kHz higher than pure shot noise. Using these values we calculate a CNR of $0.48$ for the 1 ms data. From here we determine a dwell-time of 10 ms is sufficient to achieve CNR > 1 for the 110 nm displacements. This is confirmed in Fig \ref{fig:displacement} (red) where there is almost no overlap between the signal levels at different displacements.

%(Increasing dwell-time to 10 ms reduced uncertainty $\frac{22 kHz}{\sqrt{10}} = 6.9 kHz$, which is about half the contrast of 15 kHz).
%Naturally, resolving higher vibration frequency based on SPV with faint returning optical signal is challenging due to the intrinsic trade off in detection bandwidth against dwell-time. Here, we showcase a scenario of resolve the vibration frequency at the Nyquist limit. 

We next consider the case of a harmonically vibrating target, where the motion is characterized by repeated displacements with a defined amplitude and frequency. The sensitivity of SPV to these displacements leads to modulation in the photon counting at the vibration frequency, and with contrast proportional to the vibration amplitude. For a small amplitude vibration, the modulated photon counting time-series can be expressed as a sine wave of amplitude $A$ oscillating about a mean value $\bar{N}$
\begin{equation}
    N(t) = Asin(2 \pi f t) + \bar{N},
\end{equation} 
where $f$ is the target vibration frequency. The ability to resolve the time-domain dynamics depends on the CNR of each successive measurement, and how well the signal is sampled. When the frequency is near the Nyquist limit, or the CNR is low, the signal amplitude cannot be directly measured from the time-series. In this case, a statistical approach can be used to characterize the time-domain signal. The
standard deviation of the measured signal $\sigma_m$ is the sum of the root-mean-square of the sine wave signal $\sigma_s$ and the photon counting noise $\sqrt{\bar{N}}$.
\begin{equation}
\sigma_{m} = \sqrt{ \sigma_s ^2 + \bar{N}}
\end{equation}
Recognizing this, a viable metric for characterizing the time-domain signal amplitude is to divide the measured standard deviation by the intrinsic photon counting noise to arrive at $\sigma_m/\sqrt{\bar{N}}$. This scales linearly with the vibrating signal amplitude, and is equal to 1 when the measured standard deviation is only from the shot noise. In many cases, it is not necessary to precisely resolve the dynamics of the time-domain signal. The vibration frequency and amplitude information can be obtained by identifying a peak in the frequency-domain representation of the signal, obtained by applying fast-fourier-transform (FFT) to the photon counting time-series.

We test low flux vibration sensing, and the amplitude response of SPV, by driving the PZT with a 150 Hz near sine wave signal at displacement amplitudes ranging from 0.15 $\mu m$ to 6.5 $\mu m$. The mean detection rate was fixed at about 350 kHz (0.007 detection per pulse). Data was collected for two seconds with a 1 ms dwell-time. The 150 Hz is well below the 500 Hz Nyquist frequency, allowing the time-domain waveform to be resolved and higher order harmonics to be visible in the spectrum. A zoom-in of the time-series for the largest and smallest displacements are shown in Fig \ref{fig:amplitude} (a) and (c), their FFT results in Fig \ref{fig:amplitude} (b) and (d). Larger amplitude vibration leads to high CNR between each successive measurement, directly revealing the target motion in the time-series. When the vibration amplitude is small, the signal contrast is obscured by the relative noise. In the time-domain, this noise is completely overlapped with the signal. However, after taking the FFT, the white noise is spread across the frequency domain and the signal is condensed to a narrow region, creating better contrast between the two. The ability to distinguish the signal from the white noise is given by the peak SNR, defined by the ratio between the FFT amplitude at the signal frequency and the standard deviation across other frequencies. Figure \ref{fig:amplitude} (e) shows the linear change in SNR in both frequency and time domains. The two trends are nearly identical because the total signal and noise is the same in both representations.

%\begin{equation}
  %  CNR = \frac{\sigma_{measured}}{\sqrt{\bar{N}}}
%\end{equation}

%The frequency-domain representation of the signal is created by applying fast-fourier-transform (FFT) to the time-series and normalizing by the total number of samples. The frequency domain signal is characterized by the peak SNR, which is taken as the peak height in the frequency domain divided by the standard deviation of the noise floor. It represents how the vibrating frequency can be distinguished over the interfering noise.

\begin{figure}[htbp]
\centering
\fbox{\includegraphics[width=\linewidth, height = 7 cm]{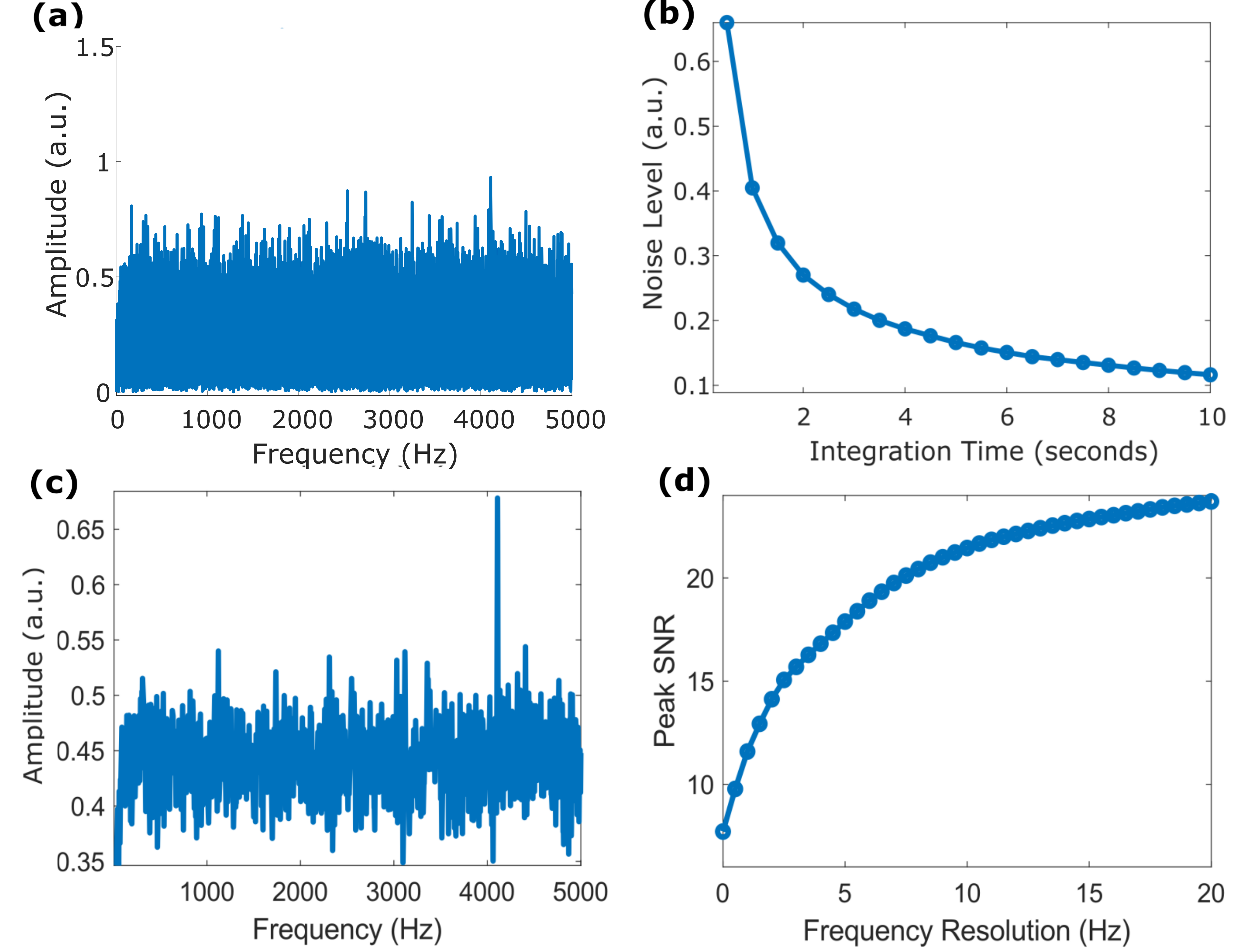}}
\caption{(a) The raw FFT results of the 4 kHz vibrating signal sampled at 10 kHz integrated over 10 seconds. (b) The frequency domain noise level (standard deviation) as a function of the total integration time. (c) FFT results in (a) after applying a 10 Hz moving mean filter, revealing the 4 kHz vibrating signal peak. (d) The peak SNR vs. the artificial frequency resolution set by the moving mean filter.}
\label{fig:highFreq}
\end{figure}

We demonstrate the limits of the current SPV bandwidth by resolving a 4 kHz vibration, with an amplitude of 2.3 $\mu$m. Here the dwell-time was shortened to 0.1 ms (10 kHz sampling rate), and the mean counting rate was 640 kHz. Higher bandwidth measurements require the use of shorter dwell-time, which naturally leads to higher photon counting noise. Additionally, the 4 kHz signal is near the Nyquist frequency and does not benefit from oversampling. Figure \ref{fig:highFreq} (a) shows the raw FFT results for a 10 second integration time. The signal peak cannot be identified due to the interference from the noise floor. However, the noise floor does reduce with a longer integration time, as shown in Fig \ref{fig:highFreq} (b). Similar to the normalized time-domain noise, the frequency domain noise level is inversely proportional to the square-root of the total integration time. This is because the photon counting shot noise is purely white noise. Figure \ref{fig:highFreq} (c) shows the results when a 10 Hz moving-mean filter is applied to the raw FFT results of the 10 second integration. The moving-mean is a common technique for smoothing data, and is sufficient post-processing to resolve the peak at 4 kHz. It is useful here to further reduce the randomly distributed noise, but comes at the cost of frequency resolution. Extending the windowing length of the filter can lead to oversmoothing and the loss of signal contrast altogether. Figure \ref{fig:highFreq} (d) shows the diminishing improvement in peak SNR as the window length is extended beyond 10 Hz, which indicates the signal linewidth to be around 10 Hz. 

%We test the performance of SPV using a PZT placed, 1 meter from the tranceiver, for precise control over the target motion. 

%The pulsed light provides inherent ranging information in its time-of-flight. Short pulses with 1 ns time-gated detection give xx cm resolution. Greater spatial resolution by scanning time-gate. Can isolate target from obscurrant. When it can't, the two surfaces can be distinguished by evaluating fourier domain of different delay settings. Two points with same mean photon number may have different ratios of signal and noise.  

%By Rayleigh criterion, the angular resolution of our 3D imager is given as  \begin{equation} \label{E0}
%\Delta \theta =2*\frac{1.22 \lambda}{D}.
%\end{equation}
%$\lambda$ is  the probe signal wavelength and $D=2*$NA$*f$ is the effective aperture of the aspheric lens limited by the numerical aperture (NA) of the SMF. With NA$\approx 0.14$ for the SMF and the focal length, $f$ = 11 mm, of aspheric lens, the angular resolution of the imager is $\approx 1.3 mrad$.
% A fiber circulator separates the outgoing signal pulses and the incoming backscattered photons with a minimum isolation ratio of 55 dB. 
 %The probe pulses, collimated to a beam diameterof 2.2 mm with very low intensity of<1 mW/mm2(picojoulesper pulse), are used to image the target behind scattering ob-stacle via a single-mode-fiber coaxial optical transceiver and a MEMS scanning mirror. 

In summary, we have demonstrated single-photon sensitive optical vibrometry based on photon counting of very faint returning signals. It samples the photon detection rate to directly measure the intensity modulated signal backscattered from a target to retrieve the vibration information. Capable of operating with less than 0.01 average detected photons per pulse, the SPV technique has shown to detect vibration displacements as small as 110 nm and resolve frequencies up to 4 kHz. The measurement bandwidth of SPV can be increased using higher repetition pulses or a more efficient detector, possibly allowing for single-photon sensitive vibrometry up to ultrasonic frequency, useful for optical elastography \cite{2020LSA.....9...58L}, acousto-optic sensing \cite{2020CmPhy...3....5D} and single photon imaging \cite{2020NatCo..11..921R}.

\section*{Disclosures.} 
The authors declare no conflicts of interest.

\bibliography{sample}
\bibliographyfullrefs{sample}

\end{document}